\documentclass[preprint,aps,showpacs,prb,amsmath,amssymb]{revtex4}
\usepackage{graphicx}
\usepackage{amssymb}
\usepackage{dcolumn}
\usepackage{amsmath}
\usepackage{bm}
\usepackage{colordvi}
\usepackage{mathrsfs}
\makeatletter

\newcommand{\Rmnum}[1]{\expandafter\@slowromancap\romannumeral #1@}
\makeatother

\begin{document}   

\title{Localized states in a semiconductor quantum ring with a tangent wire}
\author{F. Yang}
\affiliation{Hefei National Laboratory for Physical Sciences at
Microscale and Department of Physics,
University of Science and Technology of China, Hefei,
Anhui, 230026, China}

\author{M. W. Wu}
\thanks{Author to whom correspondence should be addressed}
\email{mwwu@ustc.edu.cn.}
\affiliation{Hefei National Laboratory for Physical Sciences at
Microscale and Department of Physics, University of Science and
Technology of China, Hefei, Anhui, 230026, China}

\date{\today}
\begin{abstract} 
We extend a special kind of localized state trapped at the intersection due
to the geometric confinement, first proposed in a
three-terminal-opening T-shaped structure [Euro. Phys. Lett. {\bf 55}, 539
(2001)], into a ring geometry with a tangent connection to the wire.
In this ring geometry, there exists one localized state trapped at the intersection
with energy lying inside the lowest subband. We
systematically study this localized state and the resulting Fano-type
interference due to the coupling between this localized state and 
the continuum ones. It is found that the increase of inner radius of the ring
weakens the coupling to the continuum ones and the asymmetric Fano dip fades away.  A wide
energy gap in transmission appears due to the interplay of two
types of antiresonances: the Fano-type antiresonance and the structure
antiresonance. The size of this antiresonance gap can be modulated by adjusting the
magnetic flux. Moreover, a large transmission amplitude can be obtained in the same
gap area. The strong robustness of the antiresonance gap is demonstrated and shows
the feasibility of the proposed geometry for a real application.

\end{abstract}
\pacs{73.23.Ad, 73.23.-b, 85.35.Ds,71.23.An}

\maketitle 

\section{Introduction}
 
Electrons in a T-shaped structure with three terminals opening,
i.e., three terminals extending to infinity, are classically extended. However,
numerical study for the T-shaped structure by Lin {\em et al}. showed the existence of  
a localized state trapped at the intersection.\cite{Lin} After that, Openov
presented an analytical solution of this localized state in one-dimensional
T-shaped quantum wires.\cite{Openov} Moreover, he also showed the existence
of the localized state trapped at the intersection in a four-terminal-opening
cross-shaped structure. The existence of such localized state essentially 
shows the confinement effect of the geometry in quantum region. It is
    noted that the three-terminal-opening structure here is very different from
    the previously studied T-junction system,\cite{Sols,Beaumont,Feng,Shen,Tong} where
    only two terminals are open and the structure confinement is more like a
    kind of cavity confinement.\cite{Beaumont} 
Very recently, Xu {\em et al}. investigated the localized state in the
three-terminal-opening T-shaped graphene
nanoribbons.\cite{Xu} As reported, the existence of the   
localized state due to the T-shaped confinement provides the discrete channel to
interfere with the directly propagating channels since the localized state embeds
in the continuum. This typical interference, known as the Fano-type
interference,\cite{Fano,Flach}  leads to a
characteristic asymmetric line shape in the transmission. Furthermore, if one
connects two infinity terminals in the cross-shaped 
structure, to shape a ring geometry, due to the topological similarity there
should exist at least one localized state trapped at the intersection of ring
and the attached wire. However, this localized state has not yet been studied
in the literature. 
 
Ring geometries, thanks to its special topological property, have attracted
intensive
attention.\cite{Bohm,Gefen,Webb,Levy,Chandrasekhar,Mailly,Fuhrer,Casher,Joibari,Capozza,Frustaglia,Konig,Nitta1,Bergsten,Qu,Nagasawa,Nagasawaq,Berry,Nitta2,Mal,Popp,Ionicioiu,Malshukov,Frustaglia01,Hentschel,Frustaglia04} 
For example, electrons confined to a closed path pierced by a magnetic flux, 
manifest a phase coherence phenomenon: the Aharonov-Bohm (AB) 
effect,\cite{Bohm,Gefen} which has been experimentally observed both in 
metallic\cite{Webb,Levy,Chandrasekhar} and semiconducting 
rings.\cite{Mailly,Fuhrer} In addition, in an analogous effect of the AB effect:
the Aharonov-Casher effect,\cite{Casher} the phase coherence of electrons
traveling through a closed path with spin-orbit
interaction,\cite{Joibari,Capozza,Frustaglia} has been experimentally
demonstrated through a single\cite{Konig,Nitta1,Nagasawaq} and through an array of 
mesoscopic rings.\cite{Bergsten,Qu,Nagasawa} Topics, such as Berry
phase,\cite{Berry} spin-related conductance
modulation,\cite{Nitta2,Mal} spin filters\cite{Popp} and
detectors,\cite{Ionicioiu} spin rotation\cite{Malshukov} and spin switching 
mechanisms,\cite{Frustaglia01,Hentschel,Frustaglia04} have also been exploited
intensively in ring geometries. However, up till now, the above mentioned localized state in a 
ring geometry attached to the wire has not yet drawn any attention. Recent
studies\cite{Chaves,Poniedzialek,Nowak} for a ring attached to the wire 
showed the existence of bound states, which are bounded in the ring but with small
amplitude at the intersection and almost no amplitude along the wire. However, this kind of
bound state is not the localized state as we mentioned above. Moreover, these
bound states have no coupling with the propagating states due to the
symmetry.\cite{Nowak} Accordingly, no asymmetric Fano line shape appears for
this type of bound states. 

In this work, we propose a scheme that uses a ring with a tangent connection to
the wire as well as a magnetic flux threading the ring for
modulation. We show that there exist localized states
trapped at the intersection with energies lying below the each subband.
The localized state embedded in the lower subband has a coupling with the
continuum ones which leads to the Fano-type interference. We find that this
model can provide a large tunable 
energy gap in the transmission due to the interplay of the Fano-type
antiresonance and structure antiresonance. We further demonstrate that this
energy gap is very robust against Anderson disorder, making this proposal highly
feasible for a real application.         
 
This paper is organized as follows. In Sec.~{\ref{model}},
 we set up the model and lay out the formalism. In Sec.~{\ref{Results_A}}, 
we show that for the scheme in one-dimension, there exists only 
one localized state trapped at the intersection due to the geometric
    confinement. However, no Fano 
asymmetry appears in the transmission since the energy of
the localized state lies below the energy band. In Sec.~{\ref{Results_B}}, we
find that for the quasi-one-dimensional model in practice, there exist localized 
states trapped at the intersection with energies lying below the each subband.
We systematically investigate these localized states and the resulting Fano-type
interference in the transmission. We summarize in Sec.~{\ref{Summary}}.  

\section{Model and Formalism} 
\label{model}
 
 \begin{figure}[t]
  \begin{center}
    \includegraphics[width=8.6cm]{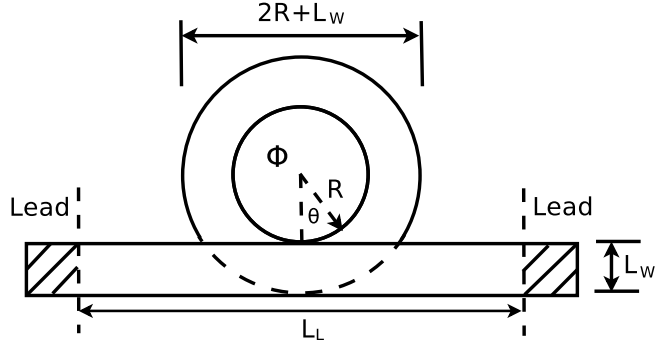}
  \end{center}
\caption{Schematic illustration of the quantum ring with a tangent connection to
  the wire. The inner and the outer rings both lie tangent to the wire
  edges. ${\Phi}$ represents the magnetic  
  flux threading the ring. ${\theta}$ denotes the azimuthal angle on the ring.}
  \label{figyw1}
\end{figure}

A schematic view of the ring structure in our study is shown in
Fig.~\ref{figyw1}, wherein a quantum ring with width $L_{W}$ and inner
    radius $R$, threaded by a magnetic flux ${\Phi}$,
    has a tangent connection to a current wire with the same width $L_{W}$
    and length $L_{L}$. The inner and the outer rings both lie tangent to the wire
  edges. ${\theta}$ denotes
  the azimuthal angle on the ring. The wire is connected to the left/right leads with the same
width $L_{W}$ through perfect ideal ohmic contacts.    

We describe the model by the
tight-binding Hamiltonian with the 
nearest-neighbor approximation, 
\begin{eqnarray}
H&=&H_{S}+H_{W}+H_{CS} +H_{L}+H_{R}+H_{CL}
\end{eqnarray}
where $H_{S}$ is the Hamiltonian of the entire ring, 
$H_{L,R,W}$ are the Hamiltonian of the left, right leads 
and the wire except the part overlapping with the ring,
respectively. $H_{CL,CS}$ stands for the couplings between the wire and
    the leads and between the wire and the ring, respectively. 
These terms are written as 
\begin{eqnarray}
H_{\alpha}&=&\varepsilon_{0}\sum_{i_\alpha} c^\dagger_{i_\alpha}c_{i_\alpha}-t\sum_{\langle
  i_\alpha,j_\alpha\rangle}c^\dagger_{i_\alpha}c_{j_\alpha}, \    \alpha=L,R,W \\
H_{S}&=&\varepsilon_{0}\sum_{i} c^\dagger_{i}c_{i}-t\sum_{\langle
  i,j\rangle}e^{ i({\theta}_{i}-{\theta}_{j}){\Phi}/{\Phi}_{0}}c^\dagger_{i}c_{j},    \\
H_{CS}&=&-t\sum_{\langle
  i_W,j_S\rangle}(c^\dagger_{i_W}c_{j_S}+H.c.),\\
H_{CL}&=&-t\sum_{\alpha= L,R}\sum_{\langle
  i_\alpha,j_S\rangle}(c^\dagger_{i_\alpha}c_{j_S}+H.c.).
u\end{eqnarray}
The index $i$ is the site coordinate in the structure and the leads and
$\langle i,j\rangle$ denotes that the sum is restricted to the nearest
neighbors. $t={\hbar}^2/(2m^*a^2)$ describes the hopping in the structure
and the leads and between the structure and the leads. Here, $m^*$ and $a$ stand
for the effective mass and lattice constant, respectively. $\varepsilon_{0}=4t$
represents the on-site energy. ${\Phi}_{0}=h/e$ is the flux quantum. The
    localized state is studied by diagonalizing the Hamiltonian $H_{S}
+H_{W}+H_{CS}$ numerically. 
 
Within the framework of the Landauer-B\"uttiker approach,\cite{Buttiker,Datta} the
transmission amplitude is given by 
\begin{equation}
T(E)={\rm Tr}[\Gamma(E)_{L}G^r_{S}(E)\Gamma(E)_{R}G^a_{S}(E)],
\end{equation}
in which  $\Gamma_{L/R}$ represents the self-energy of the isolated ideal leads
and $G^{r/a}_{S}$ denotes the retarded/advanced Green's function for the ring
structure.\cite{Datta} $E$ is the Fermi energy in the leads. 

\section{RESULTS}

In order to show that the existence of the localized state trapped at
intersection is indeed due to the geometric confinement effect similar to the
cross-shaped structure, we first study an one-dimensional model in
Sec.~{\ref{Results_A}}, which shows that there exists only one localized state 
trapped at the intersection with energy lying below the energy band. We then
lift the energy of this localized state into the energy band by using an on-site
gate voltage\cite{Silvestrov,Recher,Tong} and obtain the Fano asymmetric line
shape in the transmission. However, for the quasi-one-dimensional model in
practice (Sec.~{\ref{Results_B}}), due to the subband effect, there exist
localized states trapped at the intersection with energies lying below the each
subband. The transmission therefore is strongly modulated by the Fano-type
interference especially at the Fano antiresonance due to the coupling between
the localized state and the continuum ones. 

\subsection{Localized states and transport properties in one-dimensional model} 
\label{Results_A}

 \begin{figure}[t]
  \begin{center}
    \includegraphics[width=9.0cm]{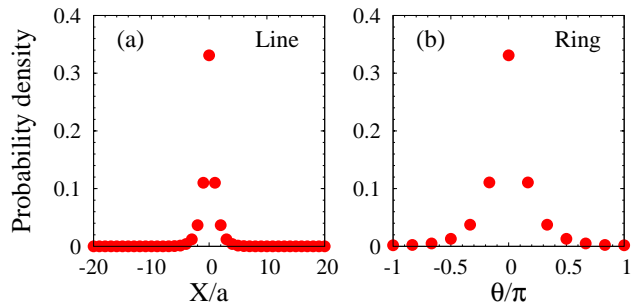}
  \end{center}
\caption{(Color online) The probability density of localized state for the one-dimensional
  scheme where the ring contains 11 lattice points. (a) The
  probability density on the wire. (b) The probability density on the ring. The
  intersection of the ring and the wire is at 0 both in (a) and (b)}   
  \label{figyw2}
\end{figure}

\begin{figure}[t]
  \begin{center}
    \includegraphics[width=11.1cm]{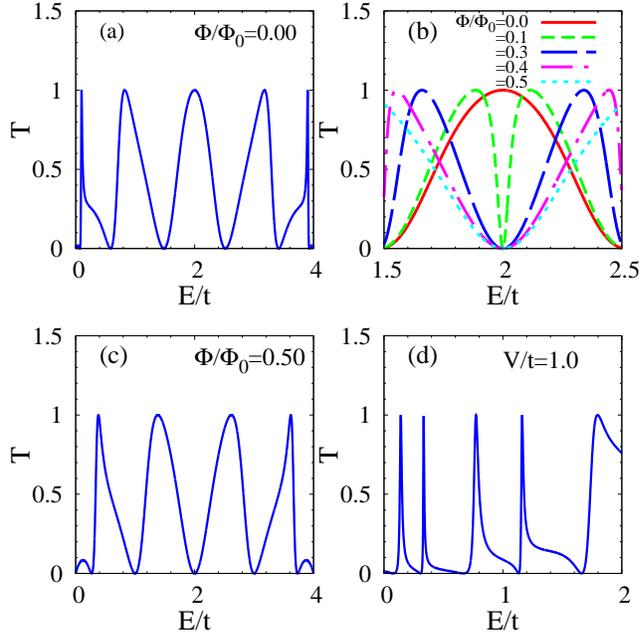}
  \end{center}
  \caption{(Color online) Transmission amplitude $T$ {\it vs} Fermi energy of the leads for
    one-dimensional scheme,
    (a) with the magnetic flux ${\Phi}/{\Phi}_{0}=0$, (b) with different
    magnetic fluxes, (c) with the magnetic flux ${\Phi}/{\Phi}_{0}=0.5$, (d) with
    an on-site gate voltage $V=1.0t$ at the intersection and ${\Phi}/{\Phi}_{0}=0$.
 }
  \label{figyw3}
\end{figure}

For the one-dimensional model ($L_{W}{\to}0$), we take the ring size
$Cl_{ring}=12a$, i.e., the ring contains 11 lattice points. The
diagonalization of the one-dimensional structure Hamiltonian $H_{S}+H_{W}+H_{CS}$ shows that
there exists only one localized state trapped at the intersection (shown in
Fig.~\ref{figyw2}) with energy lying below the energy band. Moreover, the energy of this localized state $E$ is around $-0.310t$, which is close to the energy of the localized state in one-dimensional cross-shaped structure with the same on-site and hopping energy parameters.\cite{Openov}
The transmission therefore only shows the oscillatory behavior due to the structure resonance and
antiresonance as shown in Fig.~\ref{figyw3}(a), where the transmission
amplitude is plotted against the Fermi energy $E$ in the absence of magnetic
flux. Moreover, the corresponding resonant energies in transmission, except the
two outermost resonances which are close to the energy band edges, coincide
with the eigenvalues of the isolated mesoscopic ring with the same size.\cite{Maiti}
The energies of the two outermost resonances are also very close to the
eigenvalues of the isolated mesoscopic ring. In Fig.~\ref{figyw3}(b), we plot
the transmission amplitudes of one resonance against the Fermi energy $E$ with
different magnetic fluxes $\Phi$. One observes that the resonance peak doubles
when ${\Phi}{\ne}0$ since the
magnetic flux breaks the symmetry of the clockwise and counter-clockwise
propagation on the ring, and the two resonance peaks shifted in the opposite
directions with increasing the magnetic flux. Particularly, when ${\Phi}/{\Phi}_{0}=0.5$, the
resonance and the antiresonance interchange the position in contrast to
${\Phi}/{\Phi}_{0}=0$ by comparing Figs.~\ref{figyw3}(a) and (c). When ${\Phi}/{\Phi}_{0}=1$, the
    transimission spectrum returns to the case with ${\Phi}/{\Phi}_{0}=0$ due to
    the periodicity of the magnetic flux. It is noted that we neglect the effect of the
magnetic flux on the left and right leads in the computation. This is because
that the AB phase coherence is due to the phase accumulated in a closed path rather than the 
phase accumulated on the leads. Our calculations also confirm that there is no
difference for the results with/without the magnetic flux on the leads.   

Following the idea of introducing the localized state into the energy band, we
lift the energy of the localized state by using an on-site gate voltage. We
apply a positive voltage\cite{Liao} $V=1.0t$ in the region where the probability
density of the localized state is larger than $0.1$ (as show in
Fig.~{\ref{figyw2}}, including the intersection point, two points on the line
and two points on the ring). We find that the energy of the localized state is
lifted to $E=0.54t$. The transmission amplitudes against the Fermi energy are
plotted in Fig.~\ref{figyw3}(d), where one observes that the asymmetric Fano
line shape appears at $E=0.60t$.

\subsection{Localized states and transport properties in quasi-one-dimensional model}
\label{Results_B}

For the quasi-one-dimensional model in practice, the diagonalization of the
structure Hamiltonian $H_{S}+H_{W}+H_{CS}$ shows that there exist localized states trapped at
the intersection with energies lying below the each subband. The
states which we are interested in are the ones whose energies 
lying inside the lowest subband. We find that when the inner radius of the ring $R$
is one order of magnitude smaller than the width $L_{W}$ of the wire, there
exists only one localized state trapped at the intersection with energy
lying inside the first subband. However, with increasing the inner radius of the
ring $R$ under the same width $L_{W}$, two localized states appear
in the first subband and tend
to be degenerate. Geometrically, the increase of the inner radius
separates the two positions where the ring intersects into the wire, the
intersection then gradually evolves into two small equivalent intersections of
the T-shaped structure.  The two localized states
embedded in the first subband concentrate at these two
intersections. Since we are interested in the 
localized states due to the geometric confinement, we therefore concentrate on the case with small inner radius of the
ring (one order of magnitude smaller than the width 
$L_{W}$ of the wire).   

\begin{figure}[t]
  \begin{center}
    \includegraphics[width=8.2cm]{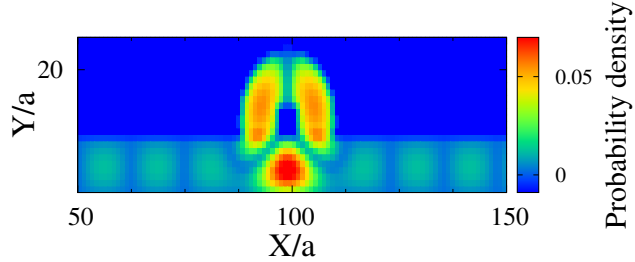}
  \end{center}
\caption{(Color online) The probability density of the quasi-bound state for the system with
  $L_{W}=18a$, $R=2a$ and ${\Phi}/{\Phi}_{0}=0.5$ with eigen-energy $E=0.0512t$.}
  \label{figyw4}
\end{figure}

In Fig.~\ref{figyw4}, we plot the probability density of the localized state for
the system with $R=2a$ and $L_{W}=18a$, and with the eigen-energy $E=0.0512t$ lying inside
the first subband. One finds from the figure that for this localized state
$\phi$ the electron indeed concentrates at the intersection, and decays exponentially away
from the intersection. Moreover, the wave-function amplitude in the wire is
small but finite, which implies that this localized state embedded in 
the first subband has a coupling ${\langle}{\phi}|H|{\Phi}{\rangle}$ with the
continuum ones $\Phi$ and therefore manifests a quasi-localized behavior. For the
system with different inner radii $R$ under the same width $L_{W}$, the
coupling  ${\langle}{\phi}|H|{\Phi}{\rangle}$ mainly comes from the region where
the localized state concentrates and the localized state trapped at 
a larger region has a stronger coupling (larger wave-function amplitude in the wire). Furthermore, we also find the adjustment of the magnetic flux has
 little influence on the position of the localized state trapped at the
 intersection in the energy spectrum.   
 
\begin{figure}[t]
  \begin{center}
    \includegraphics[width=9.6cm]{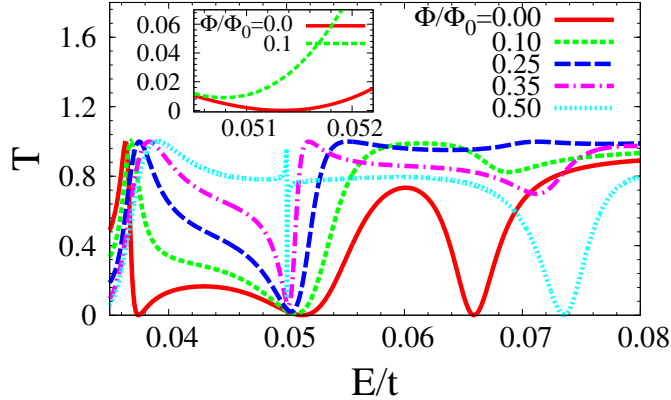}
  \end{center}
\caption{(Color online) Transmission amplitude $T$ {\it vs} Fermi energy of the leads with different
  magnetic fluxes when $L_{W}=18a$ and $R=2a$.
 }
  \label{figyw5}
\end{figure}

\begin{figure}[t]
  \begin{center}
    \includegraphics[width=9.6cm]{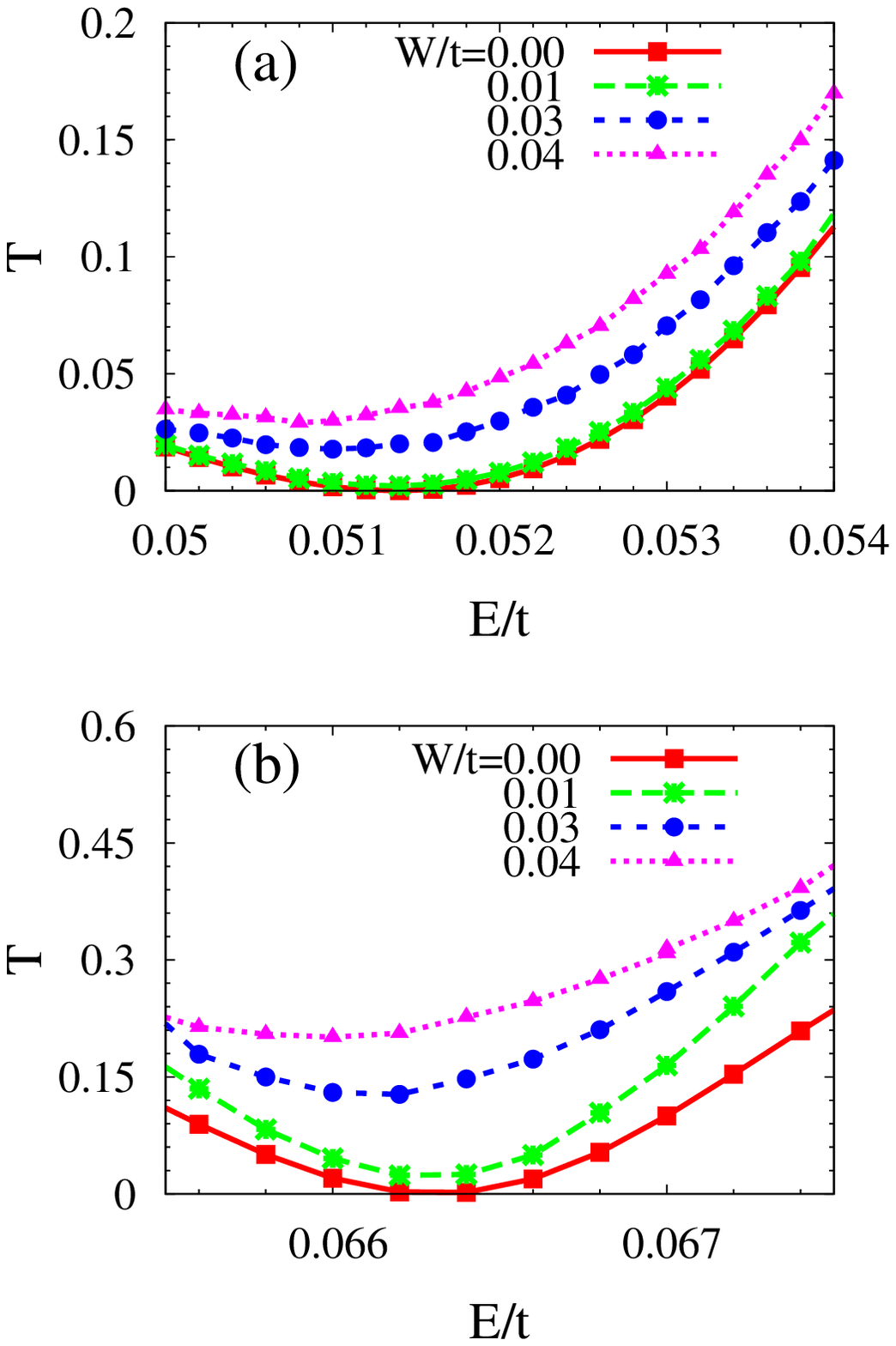}
  \end{center}
\caption{(Color online) Transmission amplitude $T$ {\it vs} Fermi energy of the leads
  with different disorder strengths $W$ when $L_{W}=18a$,
  $R=2a$ and ${\Phi}/{\Phi}_{0}=0$, (a) with both the Fano antiresonance and the
  structure antiresonance. (b) with an ordinary structure
  antiresonance.
 }
  \label{figyw6}
\end{figure}

In Fig.~\ref{figyw5}, we plot the transmission amplitudes against the Fermi
energy for the system with $R=2a$ and $L_{W}=18a$, but with different 
magnetic fluxes. The energy window we choose in the
figure lies inside the first subband. One finds that the asymmetric Fano line
shape appears at the vicinity of $E=0.0512t$. The position of the Fano line
shape coincides with the corresponding localized state obtained from the
diagonalization. One observes that the magnetic flux has little effect on the 
position of the Fano line shape but has a marked influence on its shape. This
interesting modulation is caused by the shift of the structure 
resonance which has been demonstrated in the case of one-dimensional model in
the previous subsection. When the Fano dip is close to an antiresonance dip 
(${\Phi}/{\Phi}_{0}=0$), a wide energy gap $[0.051t,0.052t]$ appears in the
transmission (see the red solid curve). Moreover, this gap can be turned off via tuning the magnetic flux. Specifically, when the Fano dip is close to a resonance peak
(${\Phi}/{\Phi}_{0}=0.5$), a sharp dip with a 
sharp peak nearby appears in the transmission and a large transmission amplitude
    $T{\approx}0.8$ is obtained for
the gap window $[0.051t,0.052t]$ (see the blue dotted curve). This feature demonstrates that the
energy gap can be turned off or on by tuning the magnetic flux. Therefore, this
model can work as a transistor with on and off features by tuning the
magnetic flux.  
    
We now show the feasibility of the above proposed model for a real application
by analyzing the robustness of the energy gap against the Anderson
disorder. In our simulation, the Anderson disorder is introduced by generating
random on-site energies at the structure sites: 
${\varepsilon_{0}}'={\varepsilon_{0}}+{\xi} W$ in Eq.~(3).
Here $W$ is the Anderson disorder strength and ${\xi}$ is a random number with a
uniform probability distribution in the range $(\mbox-1,1)$. The converged transmission
amplitudes, which are averaged over 3000 random 
configurations for $R=2a$ and $L_{W}=18a$, are plotted in Fig.~\ref{figyw6}(a)
against the Fermi energy in the vicinity of the energy
gap with different Anderson disorder strengths $W$. One finds that the
vanishingly small transmission amplitudes in the gap window $[0.051t,0.052t]$
become larger as the disorder strength $W$ increases. But even for the very
large disorder strength $W=0.04t$, the corresponding transmission amplitudes for
the gap window are still smaller than $0.04$, which is one order of magnitude
smaller than the ``on" transmission amplitudes $0.8$ as mentioned above. For
comparison, we also show the robustness of an ordinary antiresonance (with the
energy window $[0.0660,0.0666]$ and ${\Phi}/{\Phi}_{0}=0$) without a Fano
antiresonance nearby in Fig.~\ref{figyw6}(b). The transmission amplitudes in this energy window
rapidly increase with the strength of the disorder. 
Specifically, they already reach over $0.15$ at $W=0.03t$. Therefore, the
transistors only based on the structure antiresonance are very weak against the
disorder.

\begin{figure}[t]
  \begin{center}
    \includegraphics[width=8.2cm]{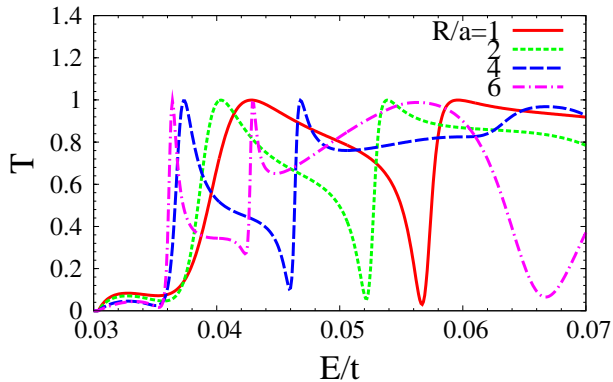}
  \end{center}
\caption{(Color online) Transmission amplitude $T$ {\it vs} Fermi energy of the leads with different
  inner radii $R$ of the ring when $L_{W}=18a$ and ${\Phi}/{\Phi}_{0}=0.35$.
 }
  \label{figyw7}
\end{figure}

In 
Fig.~\ref{figyw7}, we plot the transmission amplitudes against the Fermi energy
for the system with 
$L_{W}=18a$ and ${\Phi}/{\Phi}_{0}=0.35$ but with different inner radii $R$, one
observes that the asymmetric Fano antiresonance dip which appears in
transmission when $R=a$ (see the red solid curve) gradually fades away with
increasing the inner radius of the ring under the same width
$L_{W}=18a$. Especially, when $R=6a$ under the same width $L_{W}=18a$, this dip
in transmission is barely visible and two sharp resonance peaks appear (see the
pink chain curve). As mentioned above, under the same width $L_{W}$, the
increase of the inner radius separates the intersection into two small
ones and there appear two localized states each trapped at a small
intersection. Since the strength of the coupling
    ${\langle}{\phi}|H|{\Phi}{\rangle}$ between the localized state ${\phi}$ and
    the continuum ones ${\Phi}$ depends on the area of the localized region, these
    two localized states trapped at the small intersection therefore have weak
    couplings with the continuum ones, leading to two sharp resonance peaks in
    the transmission.

\section{SUMMARY}
\label{Summary}

In summary, we have extended a special 
kind of localized state trapped at the intersection due
to the geometric confinement, first proposed in a
three-terminal-opening T-shaped structure,\cite{Openov} into a ring
geometry with a tangent connection to the wire. This kind of the localized
states has long been overlooked in the ring geometry attached to the wire.
In this ring geometry, where we use a magnetic flux to thread the ring
for modulation, we find that when the inner radius of the ring is one order of magnitude smaller 
than the width of the attached wire, there exists one localized state trapped at
the intersection with energy lying inside the lowest subband. The Fano-type
interference due to the coupling between this localized state and the continuum ones strongly 
modulates the transmission, leading to a Fano line shape. However, the
increase of the inner radius of the ring weakens this coupling and the asymmetric
Fano dip fades away. By tuning the magnetic flux for the
structure with small inner radius of the ring, we find that a wide 
energy gap appears in the transmission when the Fano antiresonance and the
structure antiresonance are close to each other. We propose that our structure can be used as a transistor since large
transmission amplitudes can be obtained in the same energy gap region when
    the two types of the antiresonance are tuned away from each 
    other by changing the magnetic flux. We
also demonstrate the strong robustness of this energy gap against
the Anderson disorder (in contrast to an ordinary structure antiresonance). Such
features suggest that the proposed structure has great potential to work as
transistors for a real application.

\begin{acknowledgments}

This work was supported by the National Natural Science Foundation of
China under Grant No.\ 11334014, the National Basic Research Program of China
under Grant No.\ 2012CB922002, and the Strategic Priority Research Program of
the Chinese Academy of Sciences under Grant No.\ XDB01000000. One of the authors
(F.Y.) would like to thank L. Wang and T. Yu for valuable discussions.

\end{acknowledgments}

\end{document}